\documentclass[prb,aps,amsmath,amssymb,amsfonts,superscriptaddress,twocolumn]{revtex4}

\usepackage{amsmath,amssymb}
\usepackage{latexsym}
\usepackage{CJK}
\usepackage{graphicx}
\usepackage{bm}
\usepackage{color}

\newcommand{\3}{\\[3ex]}
\newcommand{\dps}{\displaystyle}
\newcommand{\om}{\iffalse}
\newcommand{\para}{\parallel}
\newcommand{\ba}{\arraycolsep 0.3 ex\begin{array}{rl}}
\newcommand{\ea}{\end{array}}
\newcommand {\pd} [2] {\frac{\partial #1}{\partial #2}}
\newcommand {\bkt} [1] {\langle #1 \rangle}

\begin{document}


\title{Electron-electron interactions in non-equilibrium bilayer graphene}


\author{Wei-Zhe Liu}
\affiliation{ICQD, Hefei National Laboratory for Physical Sciences at the Microscale, University of Science and Technology of China, Hefei 230026, Anhui, China}
\author{Allan H. MacDonald}
\affiliation{Department of Physics, The University of Texas at Austin, Austin TX 78712}
\author{Dimitrie Culcer}
\affiliation{ICQD, Hefei National Laboratory for Physical Sciences at the Microscale, University of Science and Technology of China, Hefei 230026, Anhui, China}


\begin{abstract}
Conducting steady-states of doped bilayer graphene have a non-zero sublattice pseudospin polarization. 
Electron-electron interactions renormalize this polarization even at zero temperature, when the phase space for electron-electron scattering vanishes.  We show that because of the strength of interlayer tunneling,
electron-electron interactions nevertheless have a negligible influence on the conductivity
which vanishes as the carrier number density goes to zero.  The influence of interactions 
is qualitatively weaker than in the comparable cases of single-layer graphene or topological insulators,
because the momentum-space layer pseudo spin vorticity is $2$ rather than $1$. 
Our study relies on the quantum Liouville equation in the first Born approximation with respect to the scattering potential, with electron-electron interactions taken into account self-consistently in the Hartree-Fock approximation and screening in the random phase approximation. Within this framework the result we obtain is exact.
\end{abstract}


\maketitle


\section{Introduction}




The unique properties of graphene, the first truly two-dimensional material, have spurred a flood of research on fundamental physics and on technological possibilities. \cite{Neto_Gfn_ElProp_RMP09, Castro_BBLG_ElProp_JPCM10, Nilsson_BLG_MultiLG_electron_PRB08, SDS_Gfn_RMP11, PesinAHM_Gfn_Spintron_NP12, McCann_MLGBLG_ElProp_12, McCann_BLG_ElProp_12} Monolayer graphene (MLG) has a non-Bravais honeycomb lattice structure with two triangular sublattices.  The physical consequences of its linear
$\pi$-band crossing at the Fermi level, described at low energies 
by a chiral massless Dirac $\vec{k} \cdot \vec{p}$ Hamiltonian, have been discussed at length. 
Bilayer graphene (BLG) consists of two Bernal-stacked coupled graphene monolayers, resulting in four inequivalent sites and four corresponding $\pi$-bands.  Its carriers are chiral but massive and characterized 
at low energies by a gapless parabolic spectrum displaying sublattice pseudospin-momentum locking. 
In BLG a band gap may be induced by a top gate \cite{Oostinga_BLG_GateInducedInsul_NM08, KuzmenkoCrassee_BLG_TuneGap_PRB09, Deshpande_BLG_DiracPt_APL09} or by dual gates 
which can vary the carrier density in the two layers independently. \cite{Ohta_BLG_ControlElStr_S06} 
The theory of electronic transport in this unique and tunable $\pi$-band system has been investigated extensively. \cite{Adam_BLG_Transp_PRB08, Adam_BLG_TempDepCond_PRB10, SDS_BLG_Transp_PBR10, Culcer_BLG_transp_PRB09, Hwang_BLG_inhomogeneity_temperature_PRB10, Kechedzhi_BLG_TrigWarp_PRL07, Kechedzhi_MLG_BLG_WL_EPJ07, Koshino_BLG_Deloc_PRB08, KoshinoAndo_BLG_Transp_PRB06, Min_BLG_Opt_TranspGap_PRB11, Min_MultiLG_OptCond_PRL09, Rossi_BLG_Percol_PRL11, Sinner_MLG_BLG_Dis_Cond_PRB11, Trushin_BLG_ThermCond_EPL12, Trushin_BLG_MinCond_PRB10, Cserti_BLG_MinCond_TrigWarp_PRL07, Gradinar_BLG_Strain_Lifshitz_PRB12, Li_MLG_BLG_CrgImp_SSC12, Li_BLG_insulating_subgap_conductance_NP11, Prada_BLG_QSHE_SSC11, Yuan_BLG_TLG_DisTransp_PRB10, Hatami_BLG_MgnFldCondDis_PRB11, McCannFalko_BLG_LL_QHE_PRL06, Abergel_BLG_Vly_Crnt_APL09} 
Experimental studies have focused on transport,
\cite{MiyazakiLi_BLG_unipolar_APL12, Miyazaki_BLG_Dis_PerpE_NL10, Bao_BLG_MinCond_12, Efetov_BLG_MultibandTransp_PRB11, Gorbachev_BLG_WL_PRL07, Lee_BLG_magnetotransport_NL11, Shioya_BLG_TuneNonlnrCrnt_APL12, VelascoJing_BLG_insulating_spectroscopy_NN12, Xiao_BLG_CrgImpSct_PRB10} magnetotransport, \cite{Novoselov_BLG_QHE_NP06, Henriksen_BLG_CycloRes_PRL08, Kim_BLG_SP_VP_PRL11, Sanchez-Yamagishi_BLG_Twist_QHE_PRL12, vanElferen_BLG_suspended_QHFmg_PRB12, Freitag_BLG_Susp_SpontGap_PRL12} and optics. \cite{Yang_MLG_BLG_Xtn_OptResp_PRL09, ParkLouie_Xtn_NL10, Martin_BLG_Susp_Compress_PRL10}  Recent efforts that have succeeded in manufacturing and manipulating quantum dots, \cite{Fringes_BLG_QD_PSSB11, Drocher_BLG_QD_12}  have been motivated by 
potential applications in quantum computing.

An extensive body of research has been devoted to electron-electron interactions in BLG. \cite{Barlas_BLG_neutral_nonfermi_PRB09, NandkishoreLevitov_BLG_ee_PRB10, Lemonik_BLG_GS_PRB12, Nilsson_BLG_ee_PhaseDia_PRB06,
Stauber_BLG_Fmg_JPCM08, Jung_BLG_persistent_11, Jung_BLG_PseudospinFmg_PRB11, NandkishoreLevitov_BLG_QAHE_PRB10, Min_BLG_Superfl_PRB08, Zhang_BLG_SpontSymBrk_PRB10, Zhang_BLG_SpontQHE_PRL12, Barlas_BLG_exciton_condensation_PRL10, Castro_BLG_Fmg_PRL08, Mayorov_BLG_interaction_spectrum_S11, Abergel_BLG_Correl_10, KusminskiyCampbell_BLG_ee_EPL09, KusminskiyNilsson_BLG_Compress_PRL08, Borghi_BLG_Compress_PRB10, BorghiPolini_MLGBLG_Fermi_enhancement_SSC09, Sensarma_BLG_Correl_PRB11, Vafek_BLG_ee_RG_PRB10, Killi_BBLG_LuttLiq_PRL10,
Borghi_BLG_dynamical_response_collective_PRB09, Hwang_BLG_RKKY_PRL08, Wang_BLG_Scrn_PRB07, Tse_BLG_Drude_Scrn_PRB09, Sensarma_BLG_DynScrn_PRB10, Wang_BBLG_BBLG_PRB10, NandkishoreLevitov_BLG_DynScrn_Xtn_PRL10, Gamayun_BLG_DynScrn_PRB11, TriolaRossi_BLG_Gap_Scrn_12, Abergel_BLG_Gap_Screen_12, LvWan_BLG_screening_transp_PRB10, Bena_MLG_BLG_LDOS_1Imp_PRL08, Hwang_MLG_BLG_Drag_PRB11, Rossi_Gfn_EffMed_PRB09, Hassan_BLG_pn_plasmon_PRB12} Interactions in all forms of graphene are expected to be strong when the two-dimensional 
material is surrounded by low-$\kappa$ dielectrics. 
Progress in studying interaction effects has been
improved by achieving samples with higher mobility, \cite{Morozov_MLG_BLG_IntrMob_PRL08} 
by studying suspended BLG samples that are not influenced by a substrate \cite{Feldman_BLG_Susp_BrkSym_NP09} and by breakthroughs in fabricating samples with top and back gates. \cite{Taychatanapat_BLG_DualGate_PRL10, Yan_BLG_Bolo_NN12} Interactions are expected to become more important as one approaches the charge neutrality point. \cite{Barlas_BLG_neutral_nonfermi_PRB09, NandkishoreLevitov_BLG_ee_PRB10}  In equilibrium interactions in BLG lead to competing ground states, \cite{Lemonik_BLG_GS_PRB12, Nilsson_BLG_ee_PhaseDia_PRB06} including a host of exotic states. \cite{Stauber_BLG_Fmg_JPCM08, Castro_BLG_Fmg_PRL08, Barlas_BLG_exciton_condensation_PRL10, Jung_BLG_persistent_11, Jung_BLG_PseudospinFmg_PRB11, NandkishoreLevitov_BLG_QAHE_PRB10, Min_BLG_Superfl_PRB08, Zhang_BLG_SpontSymBrk_PRB10, Zhang_BLG_SpontQHE_PRL12,Mayorov_BLG_interaction_spectrum_S11} 
Theoretical studies of electron-electron interactions in equilibrium BLG\cite{KusminskiyNilsson_BLG_Compress_PRL08, Borghi_BLG_Compress_PRB10, BorghiPolini_MLGBLG_Fermi_enhancement_SSC09, Sensarma_BLG_Correl_PRB11, KusminskiyCampbell_BLG_ee_EPL09, Abergel_BLG_Correl_10, Vafek_BLG_ee_RG_PRB10, Killi_BBLG_LuttLiq_PRL10} 
have demonstrated, among other properties, that screening and Friedel oscillations have different functional forms from MLG and 2DEGs. \cite{Hwang_BLG_RKKY_PRL08, Wang_BLG_Scrn_PRB07, Tse_BLG_Drude_Scrn_PRB09, Sensarma_BLG_DynScrn_PRB10, Wang_BBLG_BBLG_PRB10, NandkishoreLevitov_BLG_DynScrn_Xtn_PRL10, Gamayun_BLG_DynScrn_PRB11, TriolaRossi_BLG_Gap_Scrn_12, Abergel_BLG_Gap_Screen_12, Borghi_BLG_dynamical_response_collective_PRB09, LvWan_BLG_screening_transp_PRB10, Bena_MLG_BLG_LDOS_1Imp_PRL08}

The role of electron-electron interactions out of equilibrium has not yet received 
attention. Given the wealth of research on transport, it is timely to address the influence
of electron-electron interactions on the charge conductivity of BLG. 
As in single-layer graphene, charge currents in BLG lead to a net 
pseudospin polarization.  Interactions are 
therefore expected to renormalize the charge current and with it the pseudospin polarization. 
The question naturally arises of whether this polarization may be enhanced by interactions 
and produce observable effects.  It is important to understand whether the effect on the conductivity of non-equilibrium contributions to the interaction self-energy can be substantial, and whether it can be controlled using various tuning parameters such as the carrier density $n_e$.

This paper is therefore concerned with the effect of electron-electron interactions in bilayer graphene transport in the metallic regime $\varepsilon_F \tau/\hbar \gg 1$. We begin with the quantum Liouville equation for the density matrix, working in the first Born approximation with respect to momentum scattering. 
Electron-electron interactions are taken into account self-consistently using the 
non-equilibrium Hartree-Fock approximation, with screening treated in the random phase approximation. 
We determine an exact expression for the conductivity in the presence of interactions within our framework. This work is distinct from recent papers discussing other interactions in BLG transport. \cite{Hassan_BLG_pn_plasmon_PRB12, Rossi_Gfn_EffMed_PRB09, Hwang_MLG_BLG_Drag_PRB11} In addition, the mean-field effect discussed here is not related to Coulomb drag, and electron-electron scattering is not relevant to the discussion at hand, which
for simplicity assumes that the temperature $T = 0$.

We demonstrate that electron-electron interactions renormalize
the charge conductivity. The interaction effect reduces
the conductivity.  However, the effect has a very weak density dependence and 
will be difficult to distinguish experimentally from a slight increase in disorder strength, 
which is not normally known. 
Surprisingly, the interaction effect \textit{vanishes} as the carrier density $n_e$ tends towards zero
because of a subtle interplay between
the electric field, the pseudospin degree of freedom, and the electron-electron interactions
mean-field.   The effect is unexpectedly weak when the the Fermi wave vector $k_F$ is small compared to a wave vector $q_0$, introduced below, the size of which is set by the interlayer tunneling. 
In BLG this wave vector is $q_0 \approx 4$nm$^{-1}$.  Consequently even a density $\sim 10^{13}$cm$^{-2}$,
which is relatively large in experimental terms, gives a Fermi wave vector small compared to $q_0$.

A recent study of out-of-equilibrium interactions in TI showed that, for a Dirac cone, the renormalized conductivity has the same density dependence as the bare conductivity.\cite{Culcer_TI_ee_PRB11} This result is expected to apply also to MLG. In BLG, however, one would expect the renormalization to have a different density dependence given the quadratic dispersion and different functional form of the screened Coulomb potential. We find that the density dependence of the renormalization is indeed different from that of the bare conductivity, but the fractional change in the conductivity due to interactions is much weaker than in TI/MLG and vanishes at low densities.
The vanishing of the renormalization in the limit $n_e \rightarrow 0$ is explained by the fact that the pseudospin of BLG is characterized by a winding number of 2. The projection of the equilibrium 
pseudospin at ${\bm k}$ onto the pseudospin at ${\bm k}'$ has a different rotational symmetry than
the driving term due to the electric field. 
As $n_e \rightarrow 0$, the product of these terms averages to zero over the Fermi surface.
In this sense the relative weakness of this interaction effect in BLG is 
related to the smaller Fermi velocity enhancement in BLG compared to MLG. \cite{BorghiPolini_MLGBLG_Fermi_enhancement_SSC09}

The outline of this paper is as follows. In Sec.~\ref{sec:Ham}, we introduce the BLG band Hamiltonian and discuss the technical details of the kinetic equation solution.  In Sec.~\ref{sec:sct}, we calculate the scattering term in BLG in the Born approximation. In Sec.~\ref{sec:nonint} we briefly review charge transport in the absence of interactions. In Sec.~\ref{sec:int} we calculate the first order mean-field correction to the conductivity due to electron-electron interactions, then obtain an exact result to all orders. The results are discussed in Sec.~\ref{sec:disc}, while Sec.~\ref{sec:sum} summarizes our findings.

\section{Hamiltonian and Method}
\label{sec:Ham}

Our study is based on a commonly used $2$-band model for BLG: 
\begin{equation}\label{BLGHamiltonian}
    H_{0{\bm k}} = Ak^2 \left(
    \begin{array}{cccc}
        0 & e^{-2i\theta} \\
        e^{2i\theta} &0\\
    \end{array}
    \right)
    =Ak^2({\bm \sigma}\cdot\hat{\bm n}_{\bm k})
\end{equation}
where $\hat{\bm n}_{\bm k}$ is the unit vector $(\cos2\theta,\sin2\theta)^{\mathrm{T}}$ with $\theta$ the polar angle of ${\bm k}$ and $A = \hbar^2v_F^2/t_\perp$ is a material-specific constant, which determines the Fermi velocity of BLG. Here $v_F$ is the (constant) Fermi velocity of MLG and $t_\perp \approx 0.4$eV is the interlayer hopping parameter.
\cite{SDS_Gfn_RMP11} 
This model is valid at energies small compared to $t_\perp$, except that it 
neglects trigonal warping terms which become important at very low energies.
(We comment on the role of these terms in Sec.~\ref{sec:disc}.)  
It acts in a layer pseudospin space and has eigenstates which are equal weight sums of top and 
bottom layers with interlayer phase angle $ 2 \theta$.  
The Hamiltonian $H_{0{\bm k}}$ can be understood as representing a Zeeman-like interaction involving the pseudospin degree of freedom with a momentum-dependent effective magnetic field whose direction is given by 
$\hat{\bm n}_{\bm k}$. Unlike MLG, the pseudospin winds twice when 
momentum winds around the Fermi surface. 
The eigenvalues of $H_{0{\bm k}}$ are $\varepsilon_{{\bm k}\pm} = \pm Ak^2$.

The many-body Hamiltonian $\mathcal{H}$ in $2^{nd}$ quantization is
\begin{equation}
    \ba
    \displaystyle \mathcal{H} = & \displaystyle  \sum_{{\bm k}{\bm k}'ss'} (H_{{\bm k}{\bm k}'}^{ss'} c^\dag_{{\bm k}s} c_{{\bm k}'s'} +  \frac{1}{2} \, \sum_{\bm q} V_{\bm q} \, c^\dag_{{\bm k} + {\bm q}, s}c^\dag_{{\bm k}' - {\bm q}, s'}c_{{\bm k}'s'}c_{{\bm k}s}).
    \ea
\end{equation}
The one-particle matrix element $H_{{\bm k}{\bm k}'ss'}$ accounts for band structure contributions, as well as disorder and driving electric fields, discussed below.  For the matrix element $V_{\bm q} = V_{q}$ 
we use the statically screened Coulomb potential, determined here in the random phase approximation (RPA), \cite{Hwang_BLG_RKKY_PRL08} with $\epsilon_r$ the relative permittivity,
\begin{equation}
    V_q=\frac{e^2}{2\epsilon_0\epsilon_r[q + q_0g(q)]},
\end{equation}
where the constant wave vector $\displaystyle q_0=\frac{e^2}{2\pi\epsilon_0\epsilon_rA}$, and
\begin{equation}
    g(q)=\frac{1}{2k_F^2}\sqrt{4k_F^4+q^4}-\ln\bigg[\frac{k_F^2+\sqrt{k_F^4+q^4/4}}{2k_F^2}\bigg].
\end{equation}
$g(q)$ is a dimensionless function that increases monotonically from 1 to 1.755 as $q$ varies from 0 to $2k_F$, and the Fermi wave vector $k_F=\sqrt{\pi n_e}$.
Both intra-band and inter-band contributions to static screening are included in $g(q)$. For definiteness we will assume that the carrier density $n_e > 0 $. 

The effective single-particle kinetic equation for the ${\bm k}$-diagonal part of the density matrix, $f_{\bm k}$, is derived from the quantum Liouville equation in the weak momentum scattering regime exactly as in Ref.~\onlinecite{Culcer_TI_ee_PRB11}
\begin{equation}\label{Kinetic}
    \frac{df_{\bm k}}{dt} + \frac{i}{\hbar}[H_{0{\bm k}},f_{\bm k}] + \hat{J}(f_{\bm k}) = - \frac{i}{\hbar}[H_{\bm k}^E,f_{\bm k}] + \frac{i}{\hbar}[\mathcal{B}^{MF}_{\bm k},f_{\bm k}],
\end{equation}
The scattering term $\hat{J}(f_{\bm k})$ in the first Born approximation is given by
\begin{widetext}
\begin{equation}\label{Scattering}
    \ba\dps
    \hat{J}(f_{\bm k}) = \frac{n_i}{{\hbar}^2}\lim_{\eta\to0}\int\frac{d^{2}k'}{(2\pi)^2}
    |\bar{U}_{{\bm k}{\bm k'}}|^{2}\int_{0}^{\infty}dt'e^{-\eta t'}  \{e^{-iH_{0{\bm k'}}t'/\hbar}(f_{\bm k}-f_{\bm k'})e^{iH_{0{\bm k}} t'/\hbar} + e^{-iH_{0{\bm k}}t'/\hbar}(f_{\bm k}-f_{\bm k'})e^{iH_{0{\bm k}'}t'/\hbar}\},
    \ea
\end{equation}
\end{widetext}
where $n_i$ is impurity density and $\bar{U}_{{\bm k}{\bm k}'}$ the potential of a single impurity. The mean-field electron-electron interaction term is
\begin{equation}\label{EEEffective}
    \mathcal{B}^{MF}_{\bm k}(f_{\bm k})=\frac{1}{(2\pi)^2}\int dk'k'\int_{0}^{2\pi}d\gamma \, V_{{\bm k}{\bm k'}} \, f_{{\bm k}'},
\end{equation}
where $\gamma = \theta' - \theta$ is the relative angle between wave vectors ${\bm k}$ and ${\bm k}'$, $V_{{\bm k}{\bm k'}}=V_{\bm q}$, and $\theta$ and $\theta'$ are the polar angles of ${\bm k}$ and ${\bm k}'$ respectively. The one-particle Hamiltonian $H_{{\bm k}{\bm k}'}^{ss'} = H_{0{\bm k}}^{ss'} \delta_{{\bm k}{\bm k}'} + H_{E{\bm k}{\bm k}'}\delta_{ss'} + U_{{\bm k}{\bm k}'}\delta_{ss'}$, where $H_{E{\bm k}{\bm k}'}$ is the electrostatic potential due to the driving electric field ${\bm E}$, and $U_{{\bm k}{\bm k}'} $ is the \textit{total} disorder potential. 
The matrix element $\bar{U}_{{\bm k}{\bm k}'}$ of the RPA-screened Coulomb potential of a single impurity between plane waves is
\begin{equation}
  \bar{U}_{{\bm k}{\bm k'}}=\frac{Ze^2}{2\epsilon_0\epsilon_r[q+q_0g(q)]}
\end{equation}
where $Z=1$ is the ionic charge. Below we suppress the pseudospin indices $ss'$ and treat all quantities as $2 \times 2$ matrices.

We decompose $f_{\bm k}=n_{\bm k} \openone + S_{\bm k}$, with $n_{\bm k}$ a scalar part and $S_{\bm k}$ a pseudospin part, which can be expressed as $S_{\bm k}=\frac{1}{2}{\bm S}_{\bm k}\cdot{\bm \sigma}$, where the vector ${\bm S}_{\bm k}$ is real and its $z$ component is zero in equilibrium. Since the current operator is proportional to ${\bm \sigma}$, we are only interested in $S_{\bm k}$. We decompose $\hat{J}(f_{\bm k})=\hat{J}(n_{\bm k})+\hat{J}(S_{\bm k})$ and $\mathcal{B}^{MF}_{\bm k}(f_{\bm k})=\mathcal{B}^{MF}_{\bm k}(n_{\bm k})+\mathcal{B}^{MF}_{\bm k}(S_{\bm k})$, and $S_{\bm k}$ satisfies
\begin{widetext}
\begin{equation}\label{Sint}
    \ba
    \dps \frac{dS_{\bm k}}{dt}+\frac{i}{\hbar}[H_{0\bm k},S_{\bm k}]+\hat{J}(S_{\bm k}) =-\frac{i}{\hbar}[H_{\bm k}^E,S_{\bm k}]+\frac{i}{\hbar}[\mathcal{B}^{MF}_{\bm k}(S_{\bm k}),S_{\bm k}]
    \ea
\end{equation}
\end{widetext}
The interaction terms in Eq. (\ref{Sint}) can be 
included iteratively, {\em i.e.} the solution is expanded in orders of $V$, i.e. $S_{\bm k}=\sum_{n}S_{E{\bm k}}^{ee,(n)}$,
$\mathcal{B}_{k}^{MF}=\sum_{n \geqslant 0 }\mathcal{B}_{k}^{MF,(n)}$ with $\mathcal{B}^{MF,(0)}_{\bm k}=0$, and
$\mathcal{B}^{MF,(n)}_{\bm k}=\frac{1}{(2\pi)^2}\int dk'k'\int_{0}^{2\pi}d\gamma V_{{\bm k}{\bm k'}}S_{E{\bm k}}^{ee,(n-1)}$.
Here $S_{E{\bm k}} \equiv S_{E{\bm k}}^{ee,(0)}$ in the absence of electron-electron interactions. Substituting the above expansion into the kinetic equation and keeping only terms linear in the electric field (note that $\mathcal{B}_{k}^{MF}$ is linear in ${\bm E}$), we obtain the equations below for each order $n\geqslant0$:
\begin{widetext}
\begin{equation}\label{Iteration}
    \ba
    \dps\frac{dS_{E{\bm k}}^{ee,(n)}}{dt}+\frac{i}{\hbar}\big[H_{0\bm k},S_{E{\bm k}}^{ee,(n)}\big]
    +\hat{J}\big[S_{E{\bm k}}^{ee,(n)}\big]
    \dps=-\frac{i}{\hbar}\big[H_{\bm k}^{E},S_{E{\bm k}}^{ee,(n)}\big]+\frac{i}{\hbar}\big[\mathcal{B}^{MF,(n)}_{\bm k},S_{0{\bm k}} \big],
    \ea
\end{equation}
\end{widetext}
where  $S_{0{\bm k}}$ is the pseudospin-dependent part of the equilibrium density matrix, given below.

Following the method used in Ref.~\onlinecite{Culcer_TI_ee_PRB11}, we solve Eq.\ (\ref{Iteration}) for each $n$ by projecting onto directions parallel to (commuting with) and perpendicular to $H_{0{\bm k}}$, obtaining
\begin{equation}
\begin{array}{rl}
\displaystyle \frac{dS_{{\bm k}\para}}{dt}+P_{\para}\hat{J}(S_{\bm k}) = & \displaystyle D_{{\bm k}\para} \\ [1ex]
\displaystyle \frac{dS_{{\bm k}\perp}}{dt}+\frac{i}{\hbar}[H_{\bm k},S_{{\bm k}\perp}]+P_{\perp}\hat{J}(S_{\bm k}) = & \displaystyle D_{{\bm k}\perp},
\end{array}
\end{equation}
where the parallel and perpendicular components
\begin{equation}
\begin{array}{rl}
\displaystyle S_{{\bm k}\para} =  & \displaystyle (1/2)({\bm S}_{\bm k}\cdot\hat{\bm n}_{\bm k})({\bm \sigma}\cdot\hat{\bm n}_{\bm k})=
(1/2)s_{{\bm k}\para}\sigma_{{\bm k}\para} \\ [1ex]
\displaystyle S_{{\bm k}\perp} = & \displaystyle (1/2)({\bm S}_{\bm k}\cdot\hat{\bm m}_{\bm k})({\bm \sigma}\cdot\hat{\bm m}_{\bm k})=(1/2)s_{{\bm k}\perp}\sigma_{{\bm k}\perp}
\end{array}
\end{equation}
and the unit vector $\hat{\bm m}_{\bm k}=\hat{\bm z}\times\hat{\bm n}_{\bm k}$.

\section{Scattering term}
\label{sec:sct}

The scattering term does not mix the monopole ($n_{\bm k}$) and dipole ($S_{\bm k}$) components of the density matrix. Applying the decomposition ${\bm S}_{\bm k}=s_{{\bm k}\para}\hat{\bm n}_{\bm k}+s_{{\bm k}\perp}\hat{\bm m}_{\bm k}$ and ${\bm \sigma}=\sigma_{{\bm k}\para}\hat{\bm n}_{\bm k}+\sigma_{{\bm k}\perp}\hat{\bm m}_{\bm k}$, as well as $\hat{\bm n}_{\bm k'}=\cos(2\gamma)\hat{\bm n}_{\bm k}+\sin(2\gamma)\hat{\bm m}_{\bm k}$ and $\hat{\bm m}_{\bm k'}=-\sin(2\gamma)\hat{\bm n}_{\bm k}+\cos(2\gamma)\hat{\bm m}_{\bm k}$, to the expression for $\hat{J}(S_{\bm k})$, we obtain four projected terms as
\begin{equation}
\begin{array}{rl}
\displaystyle P_{\para}\hat{J}(S_{{\bm k}\para}) = & \displaystyle \frac{n_i \sigma_{{\bm k}\para}}{16\pi\hbar A}\int d{\theta}'|\bar{U}_{{\bm k}{\bm k'}}|^{2} (s_{{\bm k}\para}-s_{{\bm k'}\para}) (1 + \cos2\gamma) \\ [3ex]
\displaystyle P_{\perp}\hat{J}(S_{{\bm k}\para}) = &\displaystyle\frac{n_i \sigma_{{\bm k}\perp}}{16\pi\hbar A}\int d{\theta}'|\bar{U}_{{\bm k}{\bm k'}}|^{2} (s_{{\bm k}\para}-s_{{\bm k'}\para})\sin2\gamma
    \\ [3ex]
\displaystyle P_{\para}\hat{J}(S_{{\bm k}\perp}) = & \displaystyle \frac{n_i \sigma_{{\bm k}\para}}{16\pi\hbar A}\int d{\theta}'|\bar{U}_{{\bm k}{\bm k'}}|^{2} (s_{{\bm k}\perp}+s_{{\bm k'}\perp})\sin2\gamma
    \\ [3ex]
\displaystyle P_{\perp}\hat{J}(S_{{\bm k}\perp}) = & \displaystyle \frac{n_i \sigma_{{\bm k}\perp}}{16\pi\hbar A}\int d{\theta}'|\bar{U}_{{\bm k}{\bm k'}}|^{2}(s_{{\bm k}\perp}+s_{{\bm k'}\perp})(1 - \cos2\gamma)
\end{array}
\end{equation}

Using $q = 2k_F \sin\frac{\gamma}{2}$, we obtain a cumbersome expression for $|\bar{U}_{{\bm k}{\bm k'}}|$. We make the following Fourier expansions
\begin{equation}\label{Fourier}
\arraycolsep 0.3 ex
\begin{array}{rl}
\displaystyle |\bar{U}_{{\bm k}{\bm k'}}|^2(\gamma) = & \displaystyle \sum U_n e^{in\gamma} \\ [3ex]
\displaystyle (1 + \cos2\gamma) \, |\bar{U}_{{\bm k}{\bm k'}}|^2(\gamma) = & \displaystyle \sum W_n e^{in\gamma} \\ [3ex]
\displaystyle s_{{\bm k}\para} = & \displaystyle \sum s_{k\para n}e^{in\theta}.
\end{array}
\end{equation}
The parallel projection of the scattering term
\begin{equation}
    P_{\para}\hat{J}(S_{{\bm k}\para}) = \frac{n_i}{8\hbar A}\sum_n (W_0-W_n)s_{k\para n}e^{in\theta}\sigma_{{\bm k}\para},
\end{equation}
where $W_{-n}=W_{n}$ since $|\bar{U}_{{\bm k}{\bm k'}}|^2(\gamma)$ and $1 + \cos2\gamma$ are even functions of $\gamma$.

\section{Transport in non-interacting BLG}
\label{sec:nonint}

We briefly review transport in the absence of interactions.  Writing $S_{\bm k} = S_{0{\bm k}} + S_{E{\bm k}}$
and keeping terms to o linear order in ${\bm E}$ we obtain
\begin{equation}
    \frac{dS_{E{\bm k}}}{dt}+\frac{i}{\hbar}[H_{0\bm k},S_{E{\bm k}}]+\hat{J}(S_{E{\bm k}})=D_{E{\bm k}}
\end{equation}
where the electric-field driving term
\begin{equation}
\arraycolsep 0.3 ex
\begin{array}{rl}
\displaystyle D_{E{\bm k}} = & \displaystyle \frac{e{\bm E}}{\hbar}\cdot\pd{S_{0{\bm k}}}{\bm k}
    = \frac{1}{2} \, d_{E{\bm k}\para} \sigma_{{\bm k}\para} + \frac{1}{2} \,  d_{E{\bm k}\perp} \sigma_{{\bm k}\perp} \\ [3ex]
\displaystyle d_{E{\bm k}\para} = & \displaystyle \frac{e{\bm E}\cdot{\hat{\bm k}}}{\hbar}
    \bigg(\frac{\partial f_{0+}}{\partial k}-\frac{\partial f_{0-}}{\partial k}\bigg) \\ [3ex]
\displaystyle d_{E{\bm k}\perp} = & \displaystyle \frac{e{\bm E}\cdot\hat{\bm \theta}}{\hbar k}(f_{0+}-f_{0-}),
\end{array}
\end{equation}
in which $f_{0\pm} \equiv f_0 (\varepsilon_{{\bm k}\pm})$, with $f_0$ the Fermi-Dirac distribution function, and $S_{0{\bm k}}=(1/2)(f_{0+}-f_{0-})\sigma_{{\bm k}\para}$. We assume the temperature to be absolute zero, thus
\begin{equation}
\arraycolsep 0.3ex
\begin{array}{rl}
\displaystyle d_{E{\bm k}\para} = & \displaystyle -\frac{e{\bm E}\cdot{\hat{\bm k}}}{\hbar}\delta(k-k_F) \\ [1ex]
\displaystyle S_{{\bm k}\para} = & \displaystyle -\frac{\tau e{\bm E}\cdot\hat{\bm k}}{4\hbar}\delta(k-k_F)\sigma_{{\bm k}\para},
\end{array}
\end{equation}
where the momentum relaxation time
\begin{equation}
    \tau = \frac{8\hbar A}{n_i(W_0-W_1)}.
\end{equation}
The velocity operator is given by
\begin{equation}
{\bm v}_{\bm k} = \frac{1}{\hbar} \, \pd{H_{0{\bm k}}}{\bm k}.
\end{equation}
The expectation value of the current density operator is $\bkt{{\bm j}} = \displaystyle -eg_vg_s\int\frac{d^2k}{(2\pi)^2}\mathrm{Tr}[{\bm v}_{\bm k}S_{\bm k}]$, where Tr acts in pseudospin space, and $g_v=g_s=2$ are the valley and spin degeneracies, respectively. Substituting $S_{\bm k}=(1/2)(s_{{\bm k}\para}\sigma_{{\bm k}\para}+s_{{\bm k}\perp}\sigma_{{\bm k}\perp})$, taking ${\bm E}\,\para\,{\bm x}$, the velocity operator is $v_x=v_{x{\bm k}\para}\sigma_{{\bm k}\para}+v_{x{\bm k}\perp}\sigma_{{\bm k}\perp}$, where $v_{x{\bm k}\para}=2Ak\cos\theta/\hbar$ and $v_{x{\bm k}\perp}=-2Ak\sin\theta/\hbar$, and assuming that $\varepsilon_F \tau/\hbar \gg 1$,  it follows that the conductivity is
\begin{equation}
    \sigma_{xx}^{\mathrm{bare}} = \frac{Ae^2k_F^2\tau}{\pi\hbar^2}.
\end{equation}
The Zitterbewegung (interband coherence)  contribution to the conductivity,
plays an essential role for the minimum conductivity which survives at the charge neutrality point,\cite{Culcer_BLG_transp_PRB09, Culcer_MLG_transp_PRB08, Trushin_BLG_MinCond_PRB10, David_BLG_BS_MinCond_PRB12},
as in the MLG \cite{Culcer_MLG_transp_PRB08, Katsnelson_EPJB06} and topological insulators (TI) cases, \cite{Culcer_TI_transp_PRB10, Culcer_TI_PhysE12} but is next-to-leading order in the 
small parameter $\hbar/\varepsilon_F \tau$ and not considered here.

\section{Interaction renormalization}
\label{sec:int}

Interactions in equilibrium BLG renormalize the constant $A$ (that is, they renormalize the Fermi velocity).\cite{BorghiPolini_MLGBLG_Fermi_enhancement_SSC09} This does not make any qualitative changes to our arguments and derivation below, and for simplicity we assume henceforth that $A$ represents the renormalized $A$. In this section we will determine the mean-field interaction correction $\mathcal{B}_{\bm k}^{MF,(1)}$. From Eq.\ (\ref{Iteration}) it is evident that only the part of $\mathcal{B}_{\bm k}^{MF,(1)} \propto \sigma_{{\bm k}\perp}$ contributes to the dynamics. We abbreviate $l=k/k_F$ and
\begin{equation}\label{firstB}
    \mathcal{B}_{\bm k}^{MF,(1)}=-\frac{\tau e^3E_x}{16\pi\epsilon_0\epsilon_r\hbar}\, I_{ee}^{(1)}(l,n_e)\sin\theta\sigma_{{\bm k}\perp},
\end{equation}
where the dimensionless quantities
\begin{widetext}
\begin{equation}\label{Iee}
\arraycolsep 0.3 ex
\begin{array}{rl}
\displaystyle I_{ee}^{(1)}(l,n_e) = &\displaystyle -\int_{0}^{2\pi}\frac{d\gamma}{2\pi}\, \frac{\sqrt{\pi n_e}\sin\gamma\sin(2\gamma)}
    {\sqrt{\pi n_e(l^2+l'^2-2ll'\cos\gamma)}+q_0g(l,l',\gamma)} \\ [3ex]
\displaystyle g(l,l',\gamma) = &\displaystyle \frac{1}{2}\sqrt{4+(l^2+l'^2-2ll'\cos\gamma)^2}
    -\ln\bigg[\frac{1}{2}+\frac{1}{4}\sqrt{4+(l^2+l'^2-2ll'\cos\gamma)^2}\bigg].
    \end{array}
\end{equation}
\end{widetext}
The driving term arising from $\mathcal{B}_{\bm k}^{MF,(1)}$ contributes only to $S_{{\bm k}\perp}$, and we easily find
that 
\begin{equation}
    S_{E{\bm k}\perp}^{ee,(1)}=\frac{\tau eE_xq_0}{16\hbar k^2}I_{ee}^{(1)}(l,n_e)f_{0}
    \sin\theta\sigma_{{\bm k}\perp}.
\end{equation}
An additional correction arises from the equation
\begin{equation}
    P_{\para}\hat{J}[S_{E{\bm k}\para}^{ee,(1)}]=-P_{\para}\hat{J}[S_{E{\bm k}\perp}^{ee,(1)}],
\end{equation}
Taking this into account, the first-order correction to the diagonal conductivity is
\begin{equation}
    \sigma_{xx}^{(1)}=\frac{q_0[1+\beta(n_e)]}{4\sqrt{\pi n_e}}I_{ee}^{(1)}(n_e)\sigma_{xx}^{\mathrm{bare}}
\end{equation}
with $\beta(n_e)=(U_1-U_3)/(2U_0+2U_2-3U_1-U_3)$, in which $U_n$ is the $n$-th Fourier coefficient of $|U_{{\bm k}{\bm k'}}|^2$ as defined in Eq.\ (\ref{Fourier}), and $I^{(1)}_{ee}(n_e)=\int_{0}^{1}dl I^{(1)}_{ee}(l,n_e)$. Notice that
$\beta$ vanishes for momentum-independent (short-range) interactions.  

The angular structure of $I_{ee}^{(1)}(l,n_e)$ in Eq.\ (\ref{Iee}) can be understood by noting that the electric field driving term is responsible for the factor of $\sin\gamma$, while the factor of $\sin2\gamma$ arises from the projection of the pseudospin component parallel to $\hat{\bm k}$ onto the pseudospin component parallel to $\hat{\bm k}'$. For the massless Dirac cones of TI and MLG, where the (pseudo)spin winds around the Fermi surface only once, these terms (i.e. the electric-field driving term and the pseudospin projection) have the same rotational symmetry and reinforce each other. In BLG, the fact that the pseudospin winds twice around the Fermi surface is crucial, and makes the angular structure of this term entirely different from MLG and TI. As $n_e \rightarrow 0$, $I_{ee}^{(1)}(l,n_e)$ averages to zero over the Fermi surface. Its effect at small $n_e$ is therefore correspondingly small. In this context, it must also be noted that $q_0$ is set by $t_\perp$, the (sizable) interlayer hopping parameter, and that $q_0 \gg k_F$ even at $n = 10^{13}$cm$^{-2}$, which in transport ordinarily constitutes a large carrier density ($k_F \approx 5.5 \times 10^{8}$m$^{-1}$).





We retain only terms of linear order in the external electric field. Under these conditions, the following two equations are sufficient to obtain all higher order terms $(n>1)$,
\begin{equation}
    \ba
    \dps\frac{dS_{E{\bm k}\perp}^{ee,(n)}}{dt}+\frac{i}{\hbar}[H_{\bm k},S_{E{\bm k}\perp}^{ee,(n)}]
    =&\dps\frac{i}{\hbar}[\mathcal{B}_{\bm k}^{MF,(n)},S_{0{\bm k}}]\3
    \dps P_{\para}\hat{J}[S_{E{\bm k}\para}^{ee,(n)}]=&\dps -P_{\para}\hat{J}[S_{E{\bm k}\perp}^{ee,(n)}].
    \ea
\end{equation}
In the higher orders ($n>1$), $S_{E{\bm k}}^{ee,(n-1)}$ is fed into $\mathcal{B}^{MF,(n)}_{\bm k}$, which then determines $S_{E{\bm k}}^{ee,(n)}$, completing the self-consistent loop. Repeating the iteration, we obtain a general formula for $I_{ee}^{(n)}(n_e)$ for $n>1$, i.e.
\begin{widetext}
\begin{equation}
    \ba
    \dps I_{ee}^{(n)}(n_e)=&\dps (-\sqrt{\pi n_e})^n\bigg[\prod_{i=1}^{n-1}\int_{0}^{1}
    \frac{dl_i}{l_i}\int_{0}^{2\pi}\frac{d\gamma_i}{2\pi}\bigg]\int_{0}^{1}dl_n\int_{0}^{2\pi}\frac{d\gamma_n}{2\pi}\3
    &\dps\bigg[\prod_{i=1}^{n-1}\frac{[\cos\gamma_i\cos(2\gamma_i)+\beta(n_e) \sin\gamma_i\sin(2\gamma_i)]}{\sqrt{\pi n_e(l_i^2+l_{i+1}^2-2l_il_{i+1}\cos\gamma_i)}+q_0g(l_i,l_{i+1},\gamma_i)}\bigg]\3
    &\dps\frac{\sin\gamma_n\sin(2\gamma_n)}{\sqrt{\pi n_e(l_n^2+1-2l_n\cos\gamma_n)}+q_0g(l_n,1,\gamma_n)},
    \ea
\end{equation}
\end{widetext}
and the $n$th-order interaction correction to the conductivity is $\sigma_{xx}^{(n)}=[1+\beta(n_e)](q_0/\sqrt{16\pi n_e})^nI_{ee}^{(n)}(n_e)\sigma_{xx}^{\mathrm{bare}}$. Finally, the exact conductivity is
\begin{widetext}
\begin{equation}
    \sigma_{xx} = \sigma_{xx}^{\mathrm{bare}} \, \left\{1+[1+\beta(n_e)] \sum_{n>0}\left(\frac{q_0}{\sqrt{16\pi n_e}}\right)^nI^{(n)}_{ee}(n_e)\right\}.
\end{equation}
\end{widetext}
We refer to $\sigma_{xx}$ as the full conductivity, to distinguish it from the bare conductivity $\sigma_{xx}^{\mathrm{bare}}$. The appearance of $\beta(n_e) \ll 1$ is related to the factor of $2\gamma$ appearing in the mean-field interaction term.

\section{Discussion}
\label{sec:disc}

In equilibrium electron-electron interactions renormalize the band parameter $A$.\cite{BorghiPolini_MLGBLG_Fermi_enhancement_SSC09} 
Here we have obtained an exact result, within a self-consistent Hartree-Fock approximation, for the
influence of interactions on the conductivity of doped metallic BLG. 
Below we comment on the sign of the interaction renormalization, its size, and its density dependence.

\begin{figure}[tbp]
\bigskip
\includegraphics[width=\columnwidth]{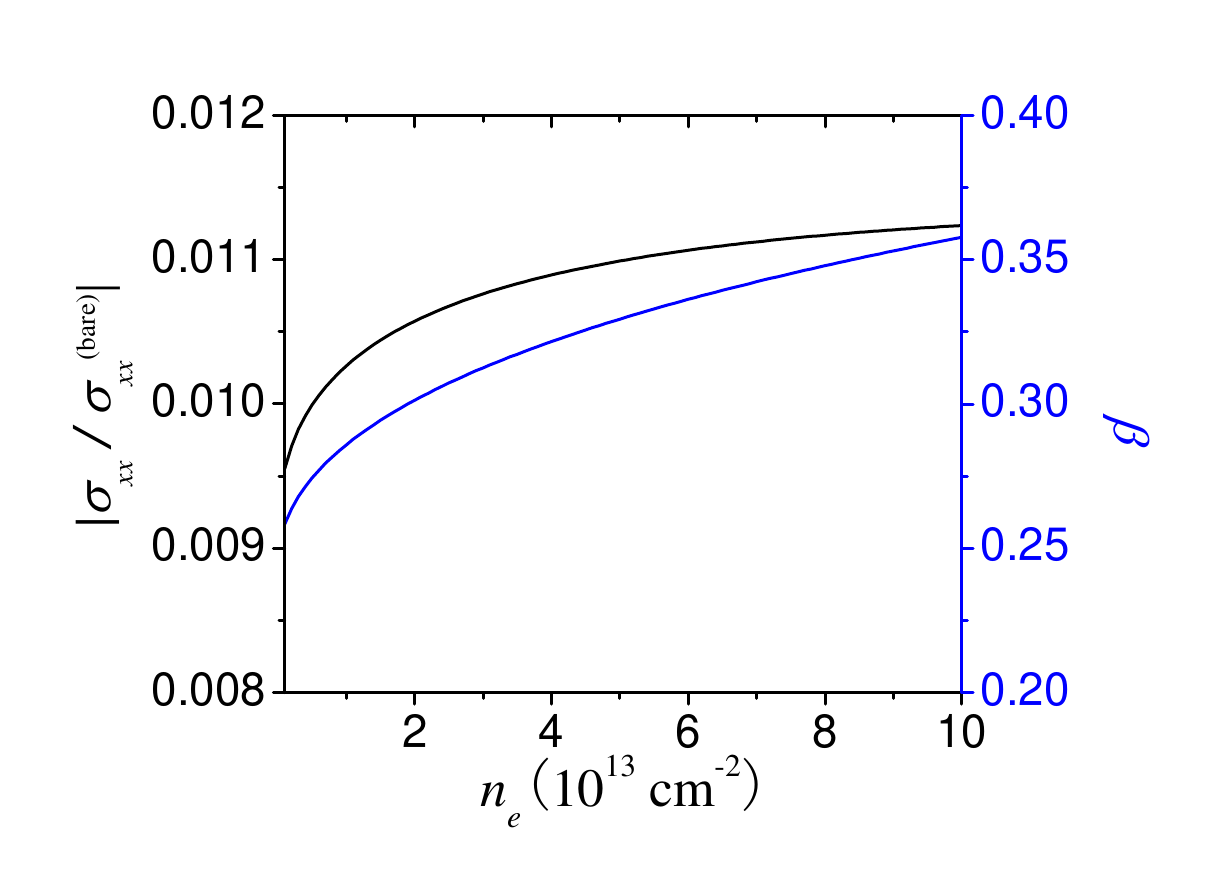}
\caption{\label{sigmavsn} Fractional change in the conductivity $\sigma_{xx}/\sigma_{xx}^{bare}$ (black), and the parameter $\beta$ (blue), as a function of the carrier density $n_e$.}
\end{figure}

We recall that a charge current in BLG 
necessarily gives rise to a steady-state pseudospin polarization. 
Consequently, $\sigma_{xx}$ may be understood by considering pseudospin dynamics on the Fermi surface. 
The renormalization reflects the interplay of pseudospin-momentum locking embodied in $H_{0{\bm k}}$ and the mean pseudospin-field $\mathcal{B}^{MF}_{\bm k}$ arising from electron-electron interactions. 
The pseudospin of one carrier on the Fermi surface at ${\bm k}$ is subject to two competing interactions. The effective field $Ak^2\hat{\bm n}_{\bm k}$ tends to align the spin with its band value. 
The mean-field $\mathcal{B}^{MF}_{\bm k}$ tends to align a pseudospin at ${\bm k}$ against the total existing pseudospin polarization. The net result is a small steady-state rotation of the pseudospin at each ${\bm k}$ away from the direction of the effective field $Ak^2\hat{\bm n}_{\bm k}$. Thus the overall effect of interactions is to align individual pseudospins in the direction opposite to that of the existing pseudospin polarization.

Since the renormalization is negative, interactions cannot cause $\sigma_{xx}$ to diverge, and
there is no possibility of a Fermi-surface instability. 
The conductivity is therefore reduced by interactions, which is reminiscent of the result of Ref.~\onlinecite{Barlas_graphene_chirality_correlation_PRL07}.
One may gain insight by further analyzing the functional form of the ratio $\sigma_{xx}/\sigma_{xx}^{bare}$, concentrating on its density dependence. Taking $A=0.71\,\mathrm{eV\cdot nm^2}$, as well as $\epsilon_r=1$ for simplicity, the wave vector $q_0=e^2/(2\pi\epsilon_0\epsilon_rA)=4.0\,\mathrm{nm^{-1}}$. As discussed before, in all realistic transport regimes $k_F=\sqrt{\pi n_e}\ll q_0$. In this low-doping regime, $\beta(n_e)$ becomes independent of $n_e$ for all $U_n\propto n_e^{1/2}$, and the conductivity simplifies to
\begin{widetext}
\begin{equation}
    \ba
    \dps{\frac{\sigma}{\sigma^{\mathrm{bare}}}}
    =&\dps 1+(1+\beta)\sum^{\infty}_{n=1}
    \left(-\frac{1}{4}\right)^n\left[
    \prod_{i=1}^{n-1}\int_{0}^{1}\frac{dl_i}{l_i}\int_{0}^{2\pi}\frac{d\gamma_i}{2\pi}\right]
    \int_{0}^{1}dl_n\int_{0}^{2\pi}\frac{d\gamma_n}{2\pi}\3
    &\dps\left[\prod_{i=1}^{n-1}\frac{\cos\gamma_i\cos(2\gamma_i)
    +\beta\sin\gamma_i\sin(2\gamma_i)}{g(l_i,l_{i+1},\gamma_i)}\right]
    \frac{\sin\gamma_n\sin(2\gamma_n)}{g(l_n,1,\gamma_n)}.
    \ea
\end{equation}
\end{widetext}
In this limit the full conductivity has almost exactly the same density dependence as the bare conductivity. The behavior at densities commonly encountered in transport is illustrated in Fig.~\ref{sigmavsn}. The small size of the renormalization makes its detection challenging. At small $n_e$, the ratio tends to zero as a result of the vanishing of angular integral appearing in $I_{ee}^{(1)}(l,n_e)$ in Eq.\ (\ref{Iee}). Steady-state expectation values are determined by the electric-field driving term, which contains a factor of ${\bm E}\cdot\hat{\bm k}$. Unlike MLG/TI, the pseudospin is not a linear function of $\parallel \hat{\bm k}$ (or $\hat{\bm \theta}$), but is characterized by a winding number of $2$. As a result of this, in BLG the interaction renormalization of the conductivity/pseudospin polarization is negligible when $k_F \ll q_0$. At large $n_e$, the behavior of $\sigma_{xx}$ is summarized by
\begin{widetext}
\begin{equation}\label{highdoping}
    \ba
    \dps\frac{\sigma}{\sigma^{\mathrm{bare}}}
    =&\dps 1-\frac{(1+\beta)q_0}{\sqrt{16\pi n_e}}
    \int_{0}^{1}dl\int_{0}^{2\pi}\frac{d\gamma}{2\pi}\frac{\sin\gamma\sin(2\gamma)}{\sqrt{l^2+1-2l\cos\gamma}}.
    \ea
\end{equation}
\end{widetext}
In this regime the ratio $\sigma_{xx}/\sigma_{xx}^{bare} \propto 1/\sqrt{n_e}$, decreases with increasing carrier density, but for this trend to be noticeable one requires $\sqrt{\pi n_e} \gg q_0$, which can never be reached in practice.

It is enlightening to compare the interaction renormalization of the conductivity in BLG with the case of TI and MLG. In TI, as in MLG, the interaction renormalization of the conductivity is density independent and again accounts for only a fraction of the total conductivity. At first sight, it seems striking that the same observation holds in BLG. Retracing the mathematical steps, the first order correction to the density matrix in TI is \cite{Culcer_TI_ee_PRB11}
\begin{equation}
    S_{E{\bm k}\perp}^{ee,(1)}=\frac{eE_xr_s\tau I_{ee}^{(1)}(l,r_s)}{16\hbar k}f_{0}\sin\theta\sigma_{{\bm k}\perp},
\end{equation}
hence $1/k$ in TI corresponds to $k_F/k^2$ in BLG, which results in approximately the same density dependence. The reason for this correspondence is that the TI Hamiltonian is $\propto k$ while the BLG Hamiltonian is $\propto k^2$, so the steady-state (pseudo)spin densities differ by a factor of $k$. At the same time, screening also differs by a factor of $k$ between the two, and the additional density dependences arising from these two factors effectively cancel out. Although the density dependence is different from TI, the correction is more complex but still weak. At very low-energies trigonal warping terms must be added to the BLG band structure,
leading to the formation of Dirac cones. In this limit, we would expect that the 
interaction correction to conductivity we discuss, would cross over to a form similar 
to that appropriate for MLG, TI's and other Dirac cone systems provided that 
this regime is not preempted by interaction-driven phase transitions to gapped states. \cite{Zhang_BLG_SpontSymBrk_PRB10, Zhang_BLG_Ordered_PRB12}     

\section{Summary}
\label{sec:sum}

We have calculated the effect of non-equilibrium interaction self-energy effects on the conductivity of metallic bilayer graphene.  Although these effects can be large in some systems, in BLG they give rise to a negative renormalization of the conductivity which is small and has a weak density dependence. This property follows from 
the large interlayer tunneling parameter in BLG, which leads to a $\pi$-band pseudo spin 
with a momentum-space winding number of $2$ that is incommensurate with the 
velocity winding number of $1$.   The corresponding effects could be larger
when a gap is opened using a bias voltage or when a magnetic field is present.

This work is supported by the National Natural Science Foundation of China under grant number 91021019. 
AHM was supported by DOE grant DE-FG03-02ER45958 and by Welch Fiundation grant No. TBF1473.
We gratefully acknowledge discussions with S.~Das Sarma.


\end{document}